\documentclass[prl,nobalancelastpage,twocolumn,superscriptaddress,showpacs,nofootinbib]{revtex4-1}
\usepackage{xcolor,amsthm,amsmath,amsfonts,graphicx,bm,amssymb,subfigure}
\usepackage{epstopdf}
\usepackage{ulem}
\usepackage{soul}

\newcommand{\MyTitle}{Engineering statistical transmutation of identical quantum particles}

\date{\today}
\begin{document}

\title{\MyTitle}

\author{Simone Barbarino}
\affiliation{SISSA, Via Bonomea 265, 34136 Trieste, Italy}
\affiliation{Institute of Theoretical Physics, Technische Universit\"at Dresden, 01062 Dresden, Germany}
\affiliation{{\rm e-mail: simone.barbarino@sns.it}}

\author{Rosario Fazio}
\affiliation{ICTP, Strada Costiera 11, 34151 Trieste, Italy}
\affiliation{NEST, Scuola Normale Superiore \& Istituto Nanoscienze-CNR, 56126 Pisa, Italy}

\author{Vlatko Vedral}
\affiliation{Centre for Quantum Technologies, National University of Singapore, 3 Science Drive 2, 117543 Singapore }
\affiliation{Clarendon Laboratory, University of Oxford, Parks Road, OX13PU Oxford, United Kingdom}

\author{Yuval Gefen}
\affiliation{Department of Condensed Matter Physics, The Weizmann Institute of Science, 76100 Rehovot, Israel}

\begin{abstract}
A fundamental pillar of quantum mechanics concerns indistinguishable quantum particles. In three dimensions they may be classified into fermions or bosons, having, respectively, antisymmetric or symmetric wave functions under particle exchange. One of numerous manifestations of this quantum statistics is the tendency of fermions (bosons) to anti-bunch (bunch). In a two-particle scattering experiment with two possible outgoing channels~\cite{hong_1987}, the probability of the two particles to arrive each at a different terminal  is enhanced (with respect to classical particles) for fermions, and reduced for bosons. Here we show that by entangling the particles with an external degree of freedom, we can 
engineer quantum statistical transmutation, e.g. causing fermions to bunch. 
Our analysis may have consequences on the observed fractional statistics of anyons, including non-Abelian statistics, with serious implications on quantum computing operations in the presence of external degrees of freedom.
\end{abstract}
\maketitle

{\it Introduction.-}The wave function of two or more indistinguishable particles must be invariant, up to a phase, if two of them are exchanged~\cite{pauli_1940,pauli_1950}. 
This requirement, together with the spin-statistics theorem, leads in three dimensions  to  the identification of fermions (with 
half-integer spin) and bosons (with integer spin), having antisymmetric or symmetric wave functions under particle exchange. 
Indistinguishability is a central pillar of quantum mechanics underlying most (if not all) facets of many particle physics. 
An elegant way to bring to light the correlations hidden in the symmetry of the wave function is through interferometry~\cite{Blanter_2000}. In a
two-particle interferometry bosons bunch together, i.e. the probability to detect  two incoming particles in the same outgoing terminal is higher than
its benchmark value for classical particles; by contrast, fermions tend to anti-bunch due to the Pauli principle: it is impossible to have coincident detections of two 
fermions with identical quantum numbers at the same terminal. The effect of quantum statistics in Hanbury-Brown-Twiss interferometric setups~\cite{Hanbury_1954,Hanbury_1956,Samuelsson_2004} has been highlighted in several fermionic-optics experiments~\cite{
Henny_1999, Oberholzer_2000, Heiblum_2007, Feve_2007, Mahe_2010, Bocquillon_2013}.

Is it possible to control and modify on-demand the quantum statistics of identical particles? Examples of emergent particles, whose quantum statistics differs 
from that of their constituents, range from Cooper pairs to  fractional statistics anyons. The latter could be viewed as fermions with attached flux lines, in the fractional quantum Hall regime~\cite{Campagnano_2012, Campagnano_2013}. 

In this Letter we show that it is possible to induce statistical transmutation in a controlled way, 
without resorting to particle-particle interaction.  The analysis presented below shows that it is possible to engineer statistical transmutation of two 
identical quantum particles by entangling them with an external quantum degree of freedom.
We study a simple setup, where quantum statistics plays a fundamental role: a two-particle scattering experiment in the 
presence of an external degree of freedom, e.g. a qubit, to which the scatterer is coupled. The build up of entanglement in the course of 
the scattering process is the underlying mechanism, allowing to engineer statistical transmutation. 

\begin{figure}
	\begin{center}
  	\includegraphics[width=\columnwidth]{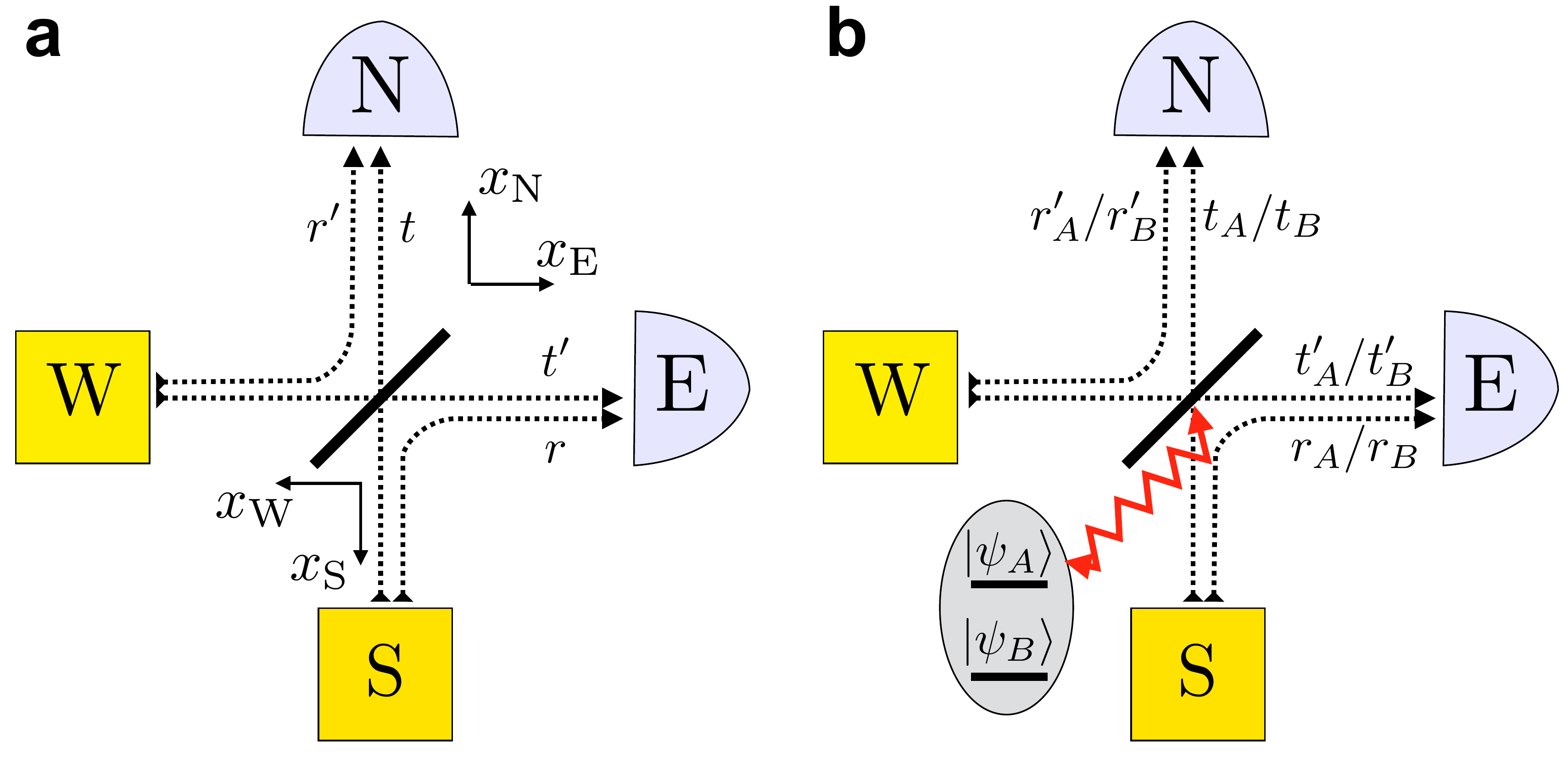}
	\end{center}
	\caption{({\bf a}) The standard Hong-Ou-Mandel interferometer: two particles emitted by the sources W and S go through 
	the scattering region modeled by a beam-splitter and they are collected by the two detectors  N and W. 
    ({\bf b}) The generalized Hong-Ou-Mandel interferometer: with respect to the standard interferometer, here the beam-splitter is coupled to an external qubit.    
	}
	\label{fig:setup}
\end{figure}

In order to analyze these effects we resort to the simple arrangement depicted in Fig.~\ref{fig:setup}{\bf a}, commonly known as a
Hong-Ou-Mandel interferometer~\cite{hong_1987, Ralley_2015}, which consists of two sources at West (W) and South (S) that emit each a single particle. 
Following the scattering event these two particles may arrive in correspondence of the detectors at North (N) and/or Est (E). 
The probability $P^{(0)}(2,0)$ ($P^{(0)}(0,2)$) would be the probability for the two scattering particles to arrive both at N (E).
The probability $P^{(0)}(1,1)$, on the contrary, would be the probability for the two scattering particles to arrive each at different detectors. 
The single-particle scattering amplitudes (hence, scattering probabilities) denote the probabilities that 
a single particle emitted from terminal $i$ ($i$=W, S) arrives in terminal $j$ ($j$=N, E). 
For classical non-interacting particles, the two-particle probabilities (e.g., $P^{(0, cl)}(2,0)$) are calculated assuming two independent scattering processes.
These classical probabilities are to be used as "benchmarks" for comparison with two-particle processes,
where quantum statistics \textit{does} play a role. For example, one finds that for fermions $P^{(0,cl)} (1,1)<P^{(0)} (1,1)$ 
while for bosons $P^{(0,cl)} (1,1)>P^{(0)} (1,1)$. Hereafter, such inequalities will be used 
to determine "femionic-like" or "bosonic-like" behavior. Statistical transmutation would mean that one type of behavior 
is transmuted to the other, \textit{i.e.} anti-bunching is transmuted into bunching and \textit{vice versa}.

Specifically, in the present analysis we consider a generalized Hong-Ou-Mandel interferometer (cf. Fig.~\ref{fig:setup}{\bf b}). Here, the scatterer (a beam splitter) is coupled to
a qubit. Consequently, additional quantum correlations between the incident particles and the qubit appear. Such
correlations can lead to non-vanishing entanglement between the scattering particles and the qubit. 
One can then project the quantum state of the qubit onto a desired direction by performing an appropriate measurement of the qubit. 
This, in turn will generate correlations between the scattered particles, that may result in non-trivial effects such as statistical transmutation. 
This means that, following such a protocol, the probability to find two fermions (bosons) in two different
outgoing terminals, i.e. anti-bunching (bunching), may be suppressed (enhanced) in comparison with the
classical benchmark, giving rise to statistical transmutation. 
Here, for the sake of clarity, we will mainly refer to fermionic particles, noting that statistical
transmutation can be obtained for both fermions and bosons. 

Beyond our engineered statistical transmutation, we note here another intriguing effect:  quantum correlations induced through the scatterer/qubit coupling may break the symmetry between the probability of collecting two particles in the N detector and collecting two particles in the E detector, $P(2,0) \neq P(0,2)$, cf. Fig.~\ref{fig:setup}{\bf b}. This symmetry breaking is a manifestation of the entanglement between the scattered particles and the qubit.

{\it Model.-} We consider a generalized Hong-Ou-Mandel interferometer, see Fig.~\ref{fig:setup}{\bf b}.
Particles are emitted from  two sources, W (West) and S (South), and are measured, after passing through a scattering region, by two 
detectors placed at N (North) and E (East). Crucial for the present discussion are the properties of the scattering region. Differently from 
the standard setup of  Fig.~\ref{fig:setup}{\bf a}, the scattering of the incoming particles comprises a two-level system coupled to a beam-splitter. 
The fermionic incoming particles are described by annihilation (creation) operators $a_\text{E}$ ($a^\dagger_\text{E}$) and $a_\text{S}$ ($a^\dagger_\text{S}$). 
The outgoing states are, in turn, described by annihilation (creation) operators  $a_\text{N}$ ($a^\dagger_\text{N}$) and $a_\text{W}$  ($a^\dagger_\text{W}$) and 
they satisfy the usual canonical anti-commutation relations. The operators of the input states and output states are not independent but are 
related by a unitary transformation $s_m$ which is just the scattering matrix~\cite{Nazarov_2009}
\begin{equation}
	\left( \begin{matrix}
	a_\text{E}\\
	a_\text{N}
	\end{matrix} \right) =
	\left(	\begin{matrix}
	r_m & t'_m\\
	t_m& r'_m
	\end{matrix}
	\right)
	\left( \begin{matrix}
	a_\text{S}\\
	a_\text{W}
	\end{matrix} \right) \,,
\end{equation}
with $r_m= \sqrt{R_m} e^{i\theta_m}$, $t'_m=\sqrt{T_m} e^{i\eta_m}$, $t_m=\sqrt{T_m}e^{i\eta_m}$, and $r'_m=-\sqrt{R_m}e^{i(2\eta_m-\theta_m)}$; 
 $\eta_m, \theta_m \in [0,2\pi)$.
The scattering matrix carries an index $m$ associated with the presence of the two-level system. When the coupling between the qubit and the beam splitter  
is non-vanishing the scattering matrix will depend on the state of the qubit. Given a basis $|\psi_m\rangle$, with $m=A,B$, for the 
state of the qubit, we have  $s_A \neq s_B$.  All these parameters do depend on the state of the qubit.

Initially the particles and the qubit are in a factorized state $|\Psi_i \rangle = |\psi_0\rangle |i\rangle$, where
$|\psi_0\rangle = \gamma_A |\psi_A\rangle +  \gamma_B |\psi_B \rangle$ is the qubit quantum state  ($|\gamma_A|^2+|\gamma_B|^2=1$)  and $|i\rangle$ is the particle state defined as
$
	|i\rangle= a^\dagger_\text{W}(x_\text{W})a^\dagger_\text{S}(x_\text{S}) |0\rangle; 
$
here the operator  
$
	a^\dagger_\ell(x_\ell) = \int_0 ^\infty \;dk_\ell \ \epsilon_\ell(k_\ell) e^{-ik_\ell x_\ell} a^\dagger_\ell(k_\ell) \label{particle_operator} 
$
with $\ell = \text{W}, \text{S}$,  creates a fermionic particle  in the source $\ell$ localized at $x_\ell$ which 
is the coordinate along the arm $\ell$ and it increases as we move along the arm toward the source. 
The operators $a_\ell(x_\ell)$ allow us to consider particles that arrive at the scatterer with a 
time delay which can be quantified in terms of  the overlap integral  
$
	|J|^2= \left |\int dk \epsilon_\text{W}(k) \epsilon^*_{\text S}(k)e^{ik(x_\text{W}-x_\text{S})}\right |^2 \,.
$ 
For $|J|^2=1$ particles are indistinguishable, while for $|J|^2=0$ particles can be considered as classical.
Following the scattering process, the final state of the system becomes 
$
 |\Psi_f\rangle =\sum_{m=A}^B \gamma_m S_m |i\rangle |\psi_m\rangle \,,
$
where $S_m = s_m^{(\text{W})} \otimes s_m^{(\text{S})}$ is the two particle scattering matrix. 
We stress here that the final state $|\Psi_f\rangle$ is generically entangled. 
As we will see this property is at the core of statistical transmutation.
 Assuming ideal detectors 
 we evaluate the probability that $n$ 
fermionic particles are revealed by the detector at N while $2-n$ particles are collected by the detector at W. Since the final state 
will be generically entangled, these probability will depend on the dynamics/measurements on the qubit. To this end, we will consider
two different protocols for the dynamics of the qubit.

{\it Protocol 1.-} Here, we evaluate the probability of particles to arrive at the detectors, independent of the state of the qubit. 
 This can be done by tracing out the qubit degrees of freedom from the density matrix of the final state
 $\rho_f=|\Psi_f\rangle \langle \Psi_f|$. This probability  
 \begin{align}
	&P(n,2-n) =\sum_{m=A}^B |\gamma_m|^2 P_m(n,2-n)  \label{PPprot1}
\end{align}
can be expressed by the weighted average of the probabilities 
$
 P_m(n,2-n)= \int dk dk' |\langle n,2-n;k,k' |S_m |i\rangle |^2
$
where 
$|2,0;k,k'\rangle = 1/\sqrt{2}\; a^\dagger_\text{N}(k) a^\dagger_\text{N}(k')|0\rangle$, $|0,2;k,k'\rangle = 1/\sqrt{2} \; 
a^\dagger_\text{E}(k) a^\dagger_\text{E}(k')|0\rangle$, 
and $|1,1;k,k'\rangle = a^\dagger_\text{N}(k) a^\dagger_\text{E}(k')|0\rangle$.  A straightforward calculation yields  
$P_m(1,1)= R^2_m+T^2_m+2R_mT_m |J|^2$ and $ P_m(2,0)=P_m(0,2)=
R_m T_m(1-|J|^2)$ (cf. Supplementary Material). 
Equation~\eqref{PPprot1} is readily understood. 
The trace over  the qubit degrees of freedom suppresses all possible quantum correlations between the particles, and the final 
result is a weighted sum over the probabilities $P_m(n,2-n)$ defined for a fixed qubit state. 
The limiting case of classical particles $P^{(cl)}(1,1)$ can be obtained, as expected, from $P_m(1,1)$ by setting $|J|^2=0$, i.e. $P^{(cl)}_m(1,1) \equiv P_m(1,1)|_{|J|^2=0}$, and employing Eq.~\eqref{PPprot1}. 
It is evident that  \textit{protocol-1} cannot lead to any statistical transmutation. For fermionic particles we still have anti-bunching, i.e.
the probability to collect the two scattered particles into two different detectors is greater for fermions than for classical particles,  
$P(1,1) > P^{(cl)}(1,1)$.

{\it Protocol 2.-}  To demonstrate how quantum correlations between the qubit and the particles may induce bunching (even for 
fermions), we now resort to projecting the qubit onto a given state, $|\psi\rangle$, following the scattering process.
We then evaluate the probability that $n$ particles are absorbed by the first detector while 
$2-n$ particles are collected by the other detector, obtaining 
\begin{equation}
	P(n,2-n;|\psi\rangle)= \frac{\int dk dk' \left| \sum_{m} \gamma_m \alpha^{(n)}_m \langle \psi 
	|\psi_m\rangle \right|^2}{\sum_{\lambda=0}^2 \int dk dk' \left| \sum_{m} \gamma_m \alpha^{(\lambda)}_m \langle \psi |	
	\psi_m\rangle \right|^2} \; ,\label{PPprot2} 
\end{equation}
with $\alpha^{(n)}_m =\langle n,2-n;k,k' |S_m |i\rangle$. 
The denominator of Eq.~\eqref{PPprot2} is a normalization factor which guarantees that $\sum_n P(n,2-n;|\psi \rangle)=1$; it 
is indeed equivalent to a post-selection procedure onto the final qubit state. The explicit calculation of Eq.~\eqref{PPprot2}  is reported in 
the Supplementary Material. {\it Protocol-2} does not have a classical counterpart.


Before proceeding, we observe that when the beam-splitter is not coupled to the qubit, i.e. $s_A=s_B \equiv s_0$, the final state $|\Psi_f\rangle$ is separable and there are no quantum correlations between particles and the qubit. In this case the two protocols are equivalent, i.e. $P(n,2-n;|\psi\rangle) =P(n,2-n)$.
For this reason, in the following,  we assume that  the beam-splitter is non trivially coupled to the qubit, i.e. $s_A \neq s_B$. As a concrete example, we
consider here the special case where $s_A$ and $s_B$ have the same amplitudes, but different phases. Under these assumptions, according to 
\textit{protocol-1} we have: $P_A(n,2-n)=P_B(n,2-n)=P(n,2-n)$ and, for classical particles, we have $P^{(cl)}(n,2-n)=\left. P(n,2-n)\right|_{|J|^2=0}$. 
Furthermore the probabilities given in Eq.~\eqref{PPprot2} can be cast in the form 
\begin{align}
	P(n,2-n;|\psi \rangle) = {\cal S}({n,|\psi \rangle})P(n,2-n)  \label{Pprot2bis}
\end{align}
with
\begin{align}
	  {\cal S}(n,|\psi \rangle) = \frac{ |\tilde \gamma_A|^2  +|\tilde \gamma_B|^2 + 2|\tilde\gamma_A||\tilde\gamma_B|\cos \varphi_{n,2-n}}
	  {|\tilde \gamma_A|^2  +|\tilde \gamma_B|^2 + 2\lambda |\tilde\gamma_A||\tilde\gamma_B| \cos \varphi_{1,1}}\,; \label{SS}
\end{align}
$
	\lambda \equiv P(1,1)+2P(2,0) \cos(\eta_B-\eta_A+\theta_A-\theta_B) 
$
and $\tilde  \gamma_m= \gamma_m \langle \psi|\psi_m\rangle \equiv |\tilde \gamma_m| e^{i \text{arg}(\tilde \gamma_m)}$; $ \varphi_{1,1}= \varphi_0+2(\eta_A-\eta_B)$, 
$\varphi_{2,0}= \varphi_0+\eta_A+\theta_A-(\eta_B+\theta_B)$, $\varphi_{0,2}= \varphi_0 +3\eta_A-\theta_A -
(3\eta_B-\theta_B)$ and  $\varphi_0=\text{arg}(\tilde\gamma_A)-\text{arg}(\tilde\gamma_B)$.
The coefficient ${\cal S}({n,|\psi \rangle})$, which depends on the number of particles detected, on the final state of the qubit, 
as well as on the phases of the scattering matrices and on the overlap integral between the incident particles,
contains all the relevant information related to the modifications of the probability distribution generated by our {\it protocol-2}. 
It is straightforward to observe that {\it protocol-2} can generate non-trivial effects when ${\cal S}({n,|\psi \rangle}) \neq 1$. We note that ${\cal S}({n,|\psi \rangle})\neq 1$ as long as the final state of the system $|\Psi_f\rangle$ is entangled and the qubit state onto which we project is in a non-trivial superposition of the states $|\psi_A\rangle$ and $|\psi_B\rangle$. 
In order to quantify whether the final state $|\Psi_f \rangle$ is entangled or not, we construct the density matrix
$
	\rho_f=|\Psi_f \rangle \langle \Psi_f|=\sum_{m,m'} \gamma_m \gamma^*_{m'} S_m |i\rangle \langle i|S^\dagger_{m'}  |\psi_m\rangle   \langle \psi_{m'}| 
$
and we then derive the reduced density matrix for the qubit by tracing out the scattered particles degrees of freedom 
\begin{align}
\rho^{red}_{f}&=
\left (
\begin{matrix}
|\gamma_A|^2 & \gamma_A \gamma^*_B \epsilon^*\\
 \gamma^*_A \gamma_B \epsilon & |\gamma_B|^2
\end{matrix}
\right);
\end{align}
here $ \langle i |S^\dagger_{A} S_B |i\rangle \equiv \epsilon$. 
A straightforward calculation  shows that the final state  $|\Psi_f\rangle$ is entangled if and only if $|J|^2 <1$ and 
$\eta_B-\eta_A+\theta_A-\theta_B \neq 0$ provided that  $\gamma_A,\,\gamma_B \neq 0$.

{\it Statistical transmutation.-} 
Before discussing our results,  we stress here again that statistical transmutation happens when the probability to find two outgoing fermions in two different terminals 
 exhibits bunching rather than the usual anti-bunching induced by the Pauli principle, i.e. $P(1,1;|\psi\rangle)$ is smaller than $P^{(cl)}(1,1)$. 
Indeed,  {\it protocol-2} allows to transmute the statistics of the incoming particles in a controlled way by properly selecting the final state $|\psi \rangle$ onto which the qubit is projected.
Importantly we observe that a necessary but insufficient requirement to obtain statistical transmutation is the presence of qubit-particles entanglement in the final state, i.e. ${\cal S}({n,|\psi \rangle})\neq 1$ (following the scattering).

 This reinforces the intuition that, in order to have transmuted statistics, one should consider composite particles (in this case 
formed by the fermions and the qubit). Note however statistical transmutation becomes possible only when ${\cal S}({1,|\psi \rangle})<P^{(cl)}(1,1)/P(1,1)$.
This last inequality  leads to 
\begin{equation}
	\frac{ |J|^2}{2P^{(cl)}(1,1)(1-|J|^2)}<\eta \label{ineq}
\end{equation}
where
\begin{equation}
	\eta=\frac{|\tilde \gamma_A| |\tilde \gamma_B|[\cos(\eta_B-\eta_A+\theta_A-\theta_B) -1] \cos \varphi_{1,1}}
	{|\tilde \gamma_A|^2  +|\tilde \gamma_B|^2 + 2 |\tilde\gamma_A||\tilde\gamma_B| \cos \varphi_{1,1}}.
\end{equation}

\begin{figure}
	\begin{center}
  	\includegraphics[width=\columnwidth]{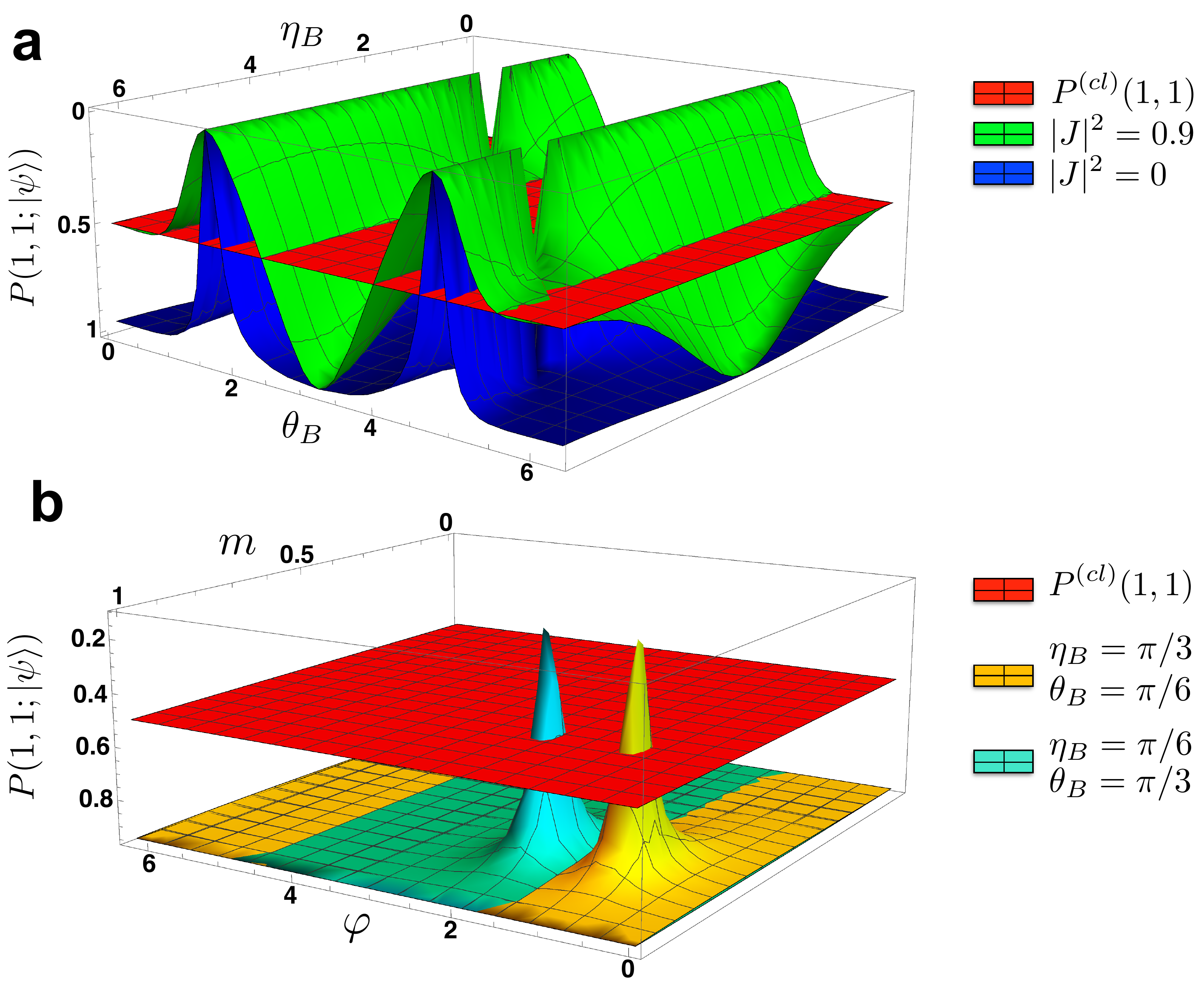}
	\end{center}
	\caption{
	The probability to find two outgoing fermions in two different terminals in a Hong-Ou-Mandel interferometer coupled to an external qubit
	can exhibit bunching rather than the usual anti-bunching dominated by the Pauli principle, i.e. $P(1,1;|\psi\rangle)<P^{(cl)}(1,1)$. 
	Statistical transmutation can be induced by selecting the final state $|\psi\rangle$ onto which the qubit is projected. 
	({\bf a}) Statistical transmution as a function of the scattering phases $\eta_B$ and $\theta_B$ with $|\psi_f \rangle =|\psi_0\rangle$. 
	({\bf b}) Statistical transmution as a function of the final state $|\psi \rangle = m |\psi_A\rangle + e^{i \varphi} \sqrt{1-m^2}  |\psi_B\rangle$ for fixed values of the scattering phases. 
	 	    }
	\label{fig:transmutation}
\end{figure}

In order to better understand the idea of statistical transmutation we consider some examples. 
In the following, we choose as initial state of the qubit $|\psi_0 \rangle =1/\sqrt{2}(|\psi_A\rangle + |\psi_B \rangle)$ and 
we set the scattering matrix amplitudes $R_{m}=T_{m}=0.5$ with $m=A,\,B$ (this choice guarantees that statistical effects are maximized, i.e. $P_m(1,1)-P_m^{(cl)}(1,1)=2R_mT_m|J|^2$ is maximum); furthermore, without loss of generality, we set $\eta_A=\theta_A=0$. 
In Fig.~\ref{fig:transmutation}{\bf a} we consider the probability  $P(1,1;|\psi\rangle)$ as a function of the phases $\eta_B$ and $\theta_B$ for a fixed final qubit state $|\psi\rangle = |\psi_0\rangle$. We observe the existence of a non-vanishing region in the parameter space $(\eta_B, \theta_B)$, where $P(1,1;|\psi\rangle)$ is smaller then $P^{(cl)}(1,1)$, as well as a region where statistical transmutation does not happen.
Moreover, from  Fig.~\ref{fig:transmutation}{\bf a} the peculiar quantum nature of {\it protocol-2} comes out. The probability $P(1,1;|\psi\rangle)$ 
with $|J|^2=0$, i.e. particles for which the Pauli principle does not apply,  still exhibits enhancement or 
suppression with respect to $P^{(cl)}(1,1)$, signaling that the qubit gives rise to additional quantum correlations beyond the Pauli principle. 
Finally, in Fig.~\ref{fig:transmutation}{\bf b}, we fix the scattering phases  $(\eta_B, \theta_B)$, and we calculate the probability $P(1,1;|\psi\rangle)$ as a function of the final state 
$|\psi\rangle =m |\psi_A\rangle + e^{i \varphi} \sqrt{1-m^2}  |\psi_B\rangle$ (here, without loss of generality, $m$ is assumed real) onto which the qubit is projected. We then observe the existence of a non-trivial manifold of final states $|\psi \rangle$, which underlines the parameter space for statistical transmutation.

{\it Unitarity breaking.-} 
Besides facilitating statistical transmutation, {\it protocol-2} allows to generate an asymmetry between the  probabilities to collect two particles in the E and in the N detector. Such an asymmetry is strictly forbidden for quantum particles, owing to the unitarity of the scattering matrix implying  $P(2,0)=P(0,2)$. 
Below we discuss under which conditions $P(2,0;|\psi\rangle)$ and $P(0,2;|\psi\rangle) $ can be different. We argue that this effect corresponds to 
an effective breaking of the unitarity of the scattering matrix. 

In order to study this effect quantitatively, we define the quantity  $\delta P(|\psi\rangle)= P(2,0;|\psi\rangle) -P(0,2;|\psi\rangle)$ which can be expressed as
\begin{equation}
\delta P(|\psi\rangle)= [ {\cal S}({2,|\psi \rangle})- {\cal S}({0,|\psi \rangle})] P(2,0) \,,
\end{equation}
and is non-vanishing when ${\cal S}({2,|\psi \rangle}) \neq {\cal S}({0,|\psi \rangle})$. In view of Eq.~\eqref{SS}, it is straightforward to see that 
$\delta P(|\psi\rangle)$ is non-vanishing as long as the final state $|\Psi_f \rangle$ is entangled. 

\begin{figure}
	\begin{center}
  	\includegraphics[width=\columnwidth]{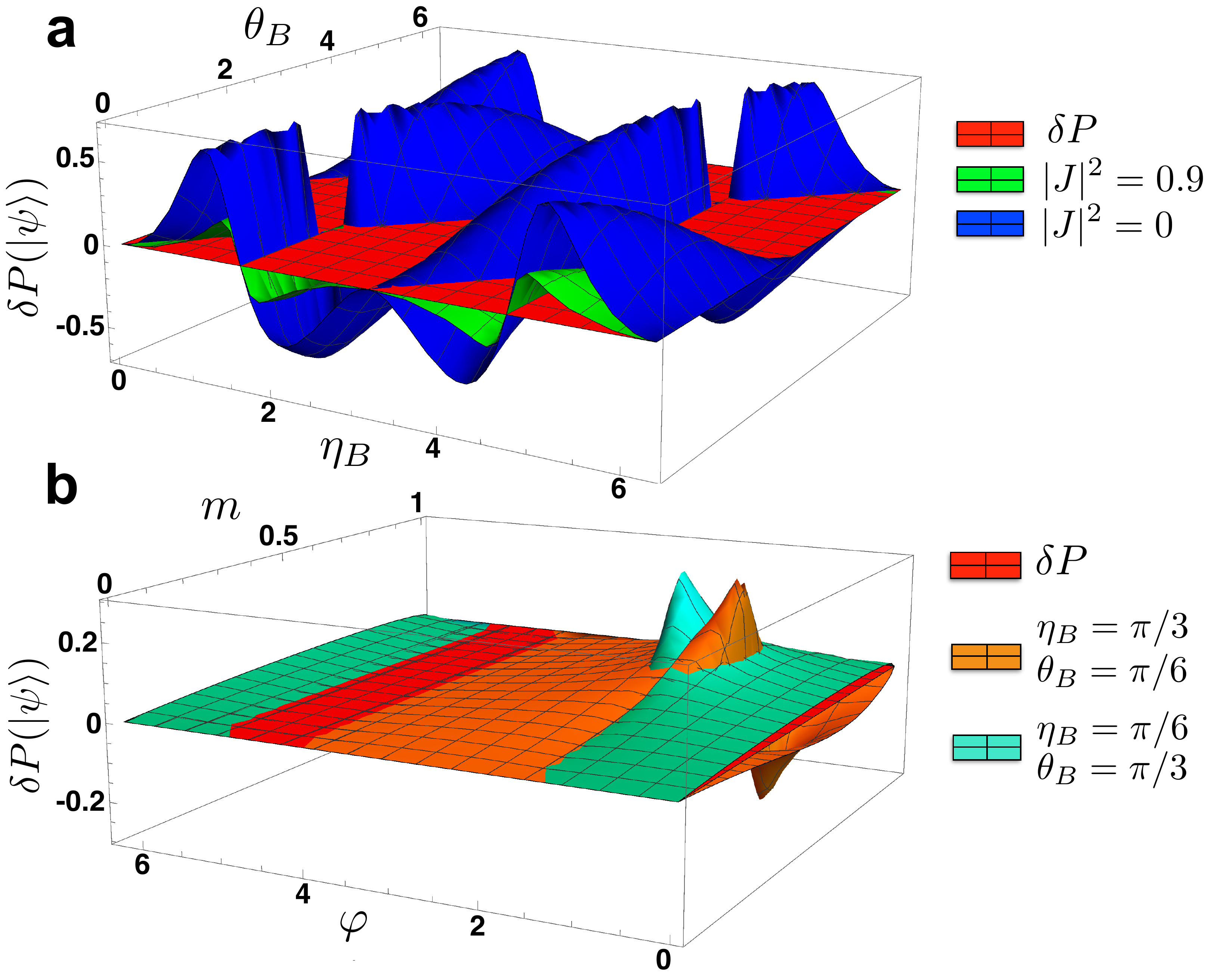}
	\end{center}
	\caption{
	The probabilities  $P(2,0;|\psi\rangle)$ and $P(0,2;|\psi\rangle)$ to find two outgoing fermions in the same terminal in a Hong-Ou-Mandel interferometer coupled to an external qubit can be different. This effect can be quantified by the quantity  $\delta P(|\psi\rangle)= P(2,0;|\psi\rangle) -P(0,2;|\psi\rangle)$.
	The asymmetry between  $P(2,0;|\psi\rangle)$ and $P(0,2;|\psi\rangle)$ can be induced by properly choosing the scattering matrix phases and the final state $|\psi\rangle$ onto the qubit is projected.
	({\bf a}) $\delta P(|\psi\rangle)$ as a function of the scattering phases $\eta_B$ and $\theta_B$ with $|\psi_f \rangle =|\psi_0\rangle$. 
	({\bf b}) $\delta P(|\psi\rangle)$ as a function of the final state $|\psi \rangle = m |\psi_A\rangle + e^{i \varphi} \sqrt{1-m^2}  |\psi_B\rangle$ for fixed values of the scattering phases. }  
	\label{fig:symmetrybreaking}
\end{figure}

In Fig.~\ref{fig:symmetrybreaking}{\bf a} we consider the asymmetry  $\delta P(|\psi\rangle)$ as a function of the phases $\eta_B$ and $\theta_B$ for a fixed final qubit state $|\psi\rangle = |\psi_0\rangle$, and we observe the existence of a non-vanishing region in parameter space where $\delta P(|\psi\rangle)$ is positive, i.e. $P(2,0;|\psi\rangle)>P(0,2;|\psi\rangle)$, or 
negative, i.e. $P(2,0;|\psi\rangle) <P(0,2;|\psi\rangle)$. 
Finally, in Fig.~\ref{fig:symmetrybreaking}{\bf b}, we fix the scattering phases  $(\eta_B, \theta_B)$ and we calculate the asymmetry $\delta P(|\psi\rangle)$ as a function of the final state 
$|\psi\rangle =m |\psi_A\rangle + e^{i \varphi} \sqrt{1-m^2}  |\psi_B\rangle$  onto which the qubit is projected. We thus obtain a non trivial manifold of final states $|\psi \rangle$ where $\delta P(|\psi\rangle)$ is non-vanishing. 

The possibility that  $\delta P(|\psi\rangle) \neq 0$  is a consequence of the additional particles-qubit correlations generated by the projection of the final state $|\Psi_f\rangle$ onto the state $|\psi\rangle$. 
Finally, it is worth noticing that for classical particles (as distinct from quantum particles for which $P(2,0)=P(0,2)$),  the probabilities $P^{(cl)}(2,0)$ and $P^{(cl)}(0,2)$ associated with a beam-splitter (even in the absence of a qubit) can be different. In this spirit, the effective breaking of the unitarity of the scattering matrix associated with our  {\it protocol-2} simulates the physics of classical particles in a scattering region.

{\it Conclusions.-}
Interaction-induced modification of quantum statistics is certainly not a new theme.  Effective attractive interactions may give rise to Cooper pairs,  with the corresponding scattering manifested by bunching.  Anyonic quasi-particles have different statistics than either fermions or bosons~\cite{Campagnano_2012, Campagnano_2013,Leinaas_1977}.
In the present work we have shown that statistical transmutation can be engineered quantum mechanically without resorting to particle-particle interaction. Specifically we have considered a generalized Hong-Ou-Mandel interferometer which allows to transmute the statistics of the scattered particles by entangling them with an external quantum degree of freedom (a qubit). Our on-demand statistics transmutation is obtained through selecting the direction in Hilbert space, onto which the qubit is \textit{a posteriori} projected. The entanglement between the scattered particles and the qubit may lead to the breaking of unitarity (still conserving probability). This can be manifest through a variety of interference setups. 
A possible experimental realization  of the Hong-Ou-Mandel interferometer may involve a quantum Hall-based setup. We tune the current in any of the interferometers to be dilute; two chiral edge modes support two beams. The scattering region is realized through a quantum point contact between the two edge states. The scatterer is then coupled electrostatically to a double quantum dot (hosting an electron) representing the qubit.

{\it Acknowledgements.-}
We acknowledge the Italia-Israel project QUANTRA, the Deutsche Forschungsgemeinschaft
(Bonn) within the network CRC TR 183 (project C01) and Grant No. RO 2247/8-1, the IMOS
Israel-Russia program, and the ISF. SB acknowledges the Institute for Theoretical Condensed Matter Physics (KIT) for support and hospitality during the completion of the work.

\onecolumngrid

\begin{center}
{\Large Supplementary Material}
\end{center}

\setcounter{equation}{0}
\setcounter{figure}{0}
\setcounter{table}{0}

\section{Protocol 1: evaluation of the probability $P(n,2-n)$}
The probabilities $P(n,2-n)$ defined according to our {\it protocol-1} are just the weighted average of the probabilities $P_m(n,2-n)$. For this reason, in the following,
we show how to calculate  the probabilities 
\begin{equation}
	P_m(n,2-n)= \int dk dk' |\langle n,2-n;k,k' |S_m |i\rangle |^2 
\end{equation}
where the particle initial state is 
$
	|i\rangle= a^\dagger_\text{W}(x_\text{W})a^\dagger_\text{S}(x_\text{S}) |0\rangle
$
with 
$
	a^\dagger_\ell(x_\ell) = \int_0 ^\infty \;dk_\ell \ \epsilon_\ell(k_\ell) e^{-ik_\ell x_\ell} a^\dagger_\ell(k_\ell)
$
and the two-particle scattering $S_m = s_m^{(\text{W})} \otimes s_m^{(\text{S})} $ is built from the single particle scattering matrix
\begin{equation}
	s_m=
	\left(	\begin{matrix}
	r_m & t'_m\\
	t_m& r'_m
	\end{matrix}
	\right) =\left(
	\begin{matrix}
	\sqrt{R_m} e^{i\theta_m}& \sqrt{T_m} e^{i\eta_m}\\
	\sqrt{T_m}e^{i\eta_m} & -\sqrt{R_m}e^{i(2\eta_m-\theta_m)} 
	\end{matrix} \right) \, ; 
\end{equation}
we recall here that we are using the basis $\{a^\dagger_\text{S} |0 \rangle, a^\dagger_\text{W}|0 \rangle \}$ ($\{a^\dagger_\text{E}|0\rangle, a^\dagger_\text{N}|0\rangle \}$) for the emitted (detected) particles. Furthermore the unitarity of $s_m$ implies:  $|r_m|^2+|t_m|^2=1$, $|t'_m|^2+|r'_m|^2=1$, and $r_mt^*_m+t'_mr'^*_m=0$
and we conveniently define $R_m=|r_m|^2=|r_m'|^2$, $T_m=|t_m|^2=|t'_m|^2$; $\eta_m, \theta_m \in [0,2\pi)$. 
Firstly we evaluate the action of the two particle scattering matrix $S_m$ onto the particle initial state 
\begin{eqnarray}
S_m|i \rangle=\int dk_\text{W} dk_\text{S} \epsilon_\text{W}(k_\text{W}) \epsilon_\text{S}(k_\text{S}) e^{-i(k_\text{W}x_\text{W}+k_\text{S}x_\text{S})} [{t'_m}a^\dagger_\text{E}(k_\text{W})+{r'_m}a^\dagger_\text{N}(k_\text{W})] [r_m a^\dagger_\text{E}(k_\text{S})+t_m a^\dagger_\text{N}(k_\text{S})] |0 \rangle \; , 
\label{finalstate}
\end{eqnarray}
then we project onto the final state $|1,1;k,k'\rangle = a^\dagger_N(k) a^\dagger_E(k')|0\rangle$ and we integrate over all possible $k$ values obtaining 
\begin{align}
P_m(1,1)&=\int dk dk' |\langle 1,1;k,k'| S_m|i \rangle|^2 =\nonumber\\
&=\int dk dk' \left|r_m{r'_m} \epsilon_{\rm W}(k) \epsilon_{\rm S}(k')e^{-ikx_{\rm W}} e^{-ik'x_{\rm S}}-t_m{t'_m}\epsilon_{\rm W}(k') \epsilon_{\rm S}(k)e^{-ik'x_{\rm W}} e^{-ikx_{\rm S}} \right|^2 =\nonumber\\
&=R_m^2+T_m^2+2R_mT_m |J|^2
\label{p11}
\end{align}
where
$
	|J|^2= \left |\int dk \epsilon_\text{W}(k) \epsilon^*_{\rm S}(k)e^{ik(x_\text{W}-x_\text{S})}\right |^2
$
is the overlap integral. 
\newline
The calculation of $P_m(2,0)$ and $P_m(0,2)$ can be carried out in a similar way. In this case, starting again from Eq.~\eqref{finalstate}, we can  
 project onto $|2,0;k,k'\rangle = 1/\sqrt{2} a^\dagger_N(k) a^\dagger_N(k')|0\rangle$ obtaining
\begin{align}
P_m(2,0)&=\int dk dk' |\langle 2,0;k,k'| S_m|i \rangle|^2 =\nonumber\\
&\int dk dk' |\langle 2,0;k,k'| S_m|i \rangle|^2 =R_m T_m \int dk dk' \left| \epsilon_{\rm W}(k) \epsilon_{\rm S}(k')e^{-ikx_{\rm W}} e^{-ik'x_{\rm S}} -\epsilon_{\rm W}(k') \epsilon_{\rm S}(k)e^{-ik'x_{\rm W}} e^{-ikx_{\rm S}}    \right|^2\nonumber \\
&=R_mT_m(1-|J|^2) \;
\label{p20}
\end{align}
similarly $P_m(0,2)$ can be obtained by projecting onto $|0,2;k,k'\rangle = 1/\sqrt{2} a^\dagger_{\rm E}(k) a^\dagger_{\rm E}(k')|0\rangle$; 
the unitarity of the scattering matrix $s_m$ implies $P_m(0,2)=P_m(2,0)$.

\section{Protocol 2: evaluation of the probability ${P}(n,2-n;|\psi\rangle)$}
We evaluate the probabilities
\begin{equation}
	P(n,2-n;|\psi\rangle)= \frac{\int dk dk' \left| \sum_{m=A}^B \gamma_m a^{(n)}_m \langle \psi 
	|\psi_m\rangle \right|^2}{\sum_{n=0}^2 \int dk dk' \left| \sum_{m=A}^B \gamma_m a^{(n)}_m \langle \psi |	
	\psi_m\rangle \right|^2} \equiv \frac{\mathcal{P}(n,2-n;|\psi\rangle)}{\sum_{n=0}^2\mathcal{P}(n,2-n;|\psi\rangle)}
	\label{P2}
\end{equation}
defined according to our {\it protocol-2}. 
Firstly, we evaluate the numerator of Eq.~\eqref{P2} for $n=0,\,1,\,2$; explicitly we obtain 
\begin{subequations}
\begin{align}
\mathcal{P}(1,1;|\psi \rangle)&= \int dk dk' \left| \gamma_A  \langle 1,1;k,k' |S_A |i\rangle \langle \psi |\psi_A\rangle +\gamma_B  \langle 1,1;k,k' |S_B |i\rangle \langle \psi |\psi_B\rangle \right|^2 =\nonumber\\
&=|\tilde \gamma_A|^2 P_A(1,1)+|\tilde \gamma_B|^2 P_B(1,1)+ 2|\tilde\gamma_A||\tilde\gamma_B|\left[R_AR_B+T_AT_B \right. \left.+(R_AT_B+R_BT_A)|J|^2\right]\cos \varphi_{1,1} 
\label{Pproj11}\\
\mathcal{P}(2,0;|\psi \rangle)&= \int dk dk' \left| \gamma_A  \langle 2,0;k,k' |S_A |i\rangle \langle \psi |\psi_A\rangle +\gamma_B  \langle 2,0;k,k' |S_B |i\rangle \langle \psi |\psi_B\rangle \right|^2 =\nonumber\\
&=|\tilde \gamma_A|^2 P_A(2,0)+|\tilde \gamma_B|^2 P_B(2,0) + 2|\tilde\gamma_A||\tilde\gamma_B| \sqrt{P_A(2,0) P_B(2,0)} \cos \varphi_{2,0} \label{Pproj20}\\
\mathcal{P}(0,2;|\psi \rangle)&= \int dk dk' \left| \gamma_A  \langle 0,2;k,k' |S_A |i\rangle \langle \psi |\psi_A\rangle +\gamma_B  \langle 0,2;k,k' |S_B |i\rangle \langle \psi |\psi_B\rangle \right|^2 =\nonumber\\
&=|\tilde \gamma_A|^2 P_A(0,2)+|\tilde \gamma_B|^2 P_B(0,2)  +  2|\tilde\gamma_A||\tilde\gamma_B| \sqrt{P_A(0,2) P_B(0,2)} \cos \varphi_{0,2} \label{Pproj02}
\end{align}
\end{subequations}
where we have introduced $\tilde  \gamma_m= \gamma_m \langle \psi|\psi_m\rangle \equiv |\tilde \gamma_m| e^{i \text{arg}(\tilde \gamma_m)}$, $m=A,B$;  
$
\varphi_0=\text{arg}(\tilde\gamma_A)-\text{arg}(\tilde\gamma_B)
$
and 
\begin{subequations}
\begin{align}
&\varphi_{1,1}= \varphi_0+2(\eta_A-\eta_B) \\
&\varphi_{2,0}= \varphi_0+\eta_A+\theta_A-(\eta_B+\theta_B)\\
&\varphi_{0,2}= \varphi_0 +3\eta_A-\theta_A -(3\eta_B-\theta_B) \,;
\end{align}
\end{subequations}
the probabilities $P_m(n,2-n)$ with $m=A,\, B$ and $n=0,\,1,\,2$ are defined in Eqs.~\eqref{p11} and~\eqref{p20}.
\newline
The denominator of Eq.~\eqref{P2} turns out to be equal to 
\begin{align}
 &\sum_{n=0}^2\mathcal{P}(n,2-n;|\psi\rangle) =|\tilde \gamma_A|^2 +|\tilde \gamma_B|^2  +  \nonumber\\
 &+2|\tilde\gamma_A||\tilde\gamma_B| \left[\left[R_AR_B+T_AT_B \right. \left.+(R_AT_B+R_BT_A)|J|^2\right]\cos \varphi_{1,1} +  \sqrt{P_A(2,0) P_B(2,0)} \left(\cos \varphi_{2,0} +  \cos \varphi_{0,2}\right) \right].
 \label{denom}
\end{align}
In the special case where $s_A$ and $s_B$ have the same amplitudes  $P_A(n,2-n)=P_B(n,2-n) \equiv P(n,2-n)$ for $n=0,1,2$. From Eq.~\eqref{denom} we obtain
\begin{align}
 &\sum_{n=0}^2\mathcal{P}(n,2-n;|\psi\rangle) =|\tilde \gamma_A|^2 +|\tilde \gamma_B|^2 +2|\tilde\gamma_A||\tilde\gamma_B| + 2\lambda |\tilde\gamma_A||\tilde\gamma_B| \cos \varphi_{1,1}
 \end{align}
where $\lambda \equiv P(1,1)+2P(2,0) \cos(\eta_B-\eta_A+\theta_A-\theta_B) $
and finally, recalling Eqs.~\eqref{Pproj11}-~\eqref{Pproj02}, we obtain $P(n,2-n;|\psi\rangle)= {\cal S}({n,|\psi \rangle }) P(n,2-n)$
with
\begin{align}
	  {\cal S}({n,|\psi \rangle})= \frac{ |\tilde \gamma_A|^2  +|\tilde \gamma_B|^2 + 2|\tilde\gamma_A||\tilde\gamma_B|\cos \varphi_{n,2-n}}
	  {|\tilde \gamma_A|^2  +|\tilde \gamma_B|^2 + 2\lambda |\tilde\gamma_A||\tilde\gamma_B| \cos \varphi_{1,1}}
\end{align}
which is the expression presented in the main text.
 
\end{document}